\documentclass[11pt,english]{iopart}

\makeatletter
\usepackage{babel}
\usepackage{hyperref}

\begin{document}

\title{The three-dimensional noncommutative Gross-Neveu model}

\author{B. Charneski and A. F. Ferrari and M. Gomes}
\address{Instituto de F\'\i sica, Universidade de S\~ao Paulo\\
Caixa Postal 66318, 05315-970, S\~ao Paulo, SP, Brazil}
\ead{bruno,alysson,mgomes@fma.if.usp.br} 

\begin{abstract}
This work is dedicated to the study of the noncommutative Gross-Neveu model. As it is known, in the canonical Weyl-Moyal approach the model is inconsistent, basically due to the separation of the amplitudes into planar and nonplanar parts. We prove that if instead a coherent basis representation is used, the model becomes renormalizable and free of the aforementioned difficulty. We also show that, although the coherent states procedure breaks Lorentz symmetry in odd dimensions, in the Gross-Neveu model this breaking can be kept under control by assuming the noncommutativity parameters to be small enough. We also make some remarks on ordering prescriptions used in the literature.
\end{abstract}


\maketitle

\section{Introduction}
\label{intro}

In the recent years much effort has been
devoted to the study of noncommutative field theories~\cite{reviews}. One important
outcome of these investigations is that, for the case of canonical
noncommutativity, the use of the Weyl-Moyal correspondence leads to strong  nonlocal effects,
which put severe restrictions on the form of the allowed models. In fact, it has been found that part of
the ultraviolet divergences of the commutative models are transmuted
into infrared ones. Whenever they are stronger than logarithmic, these
divergences, called ultraviolet/infrared (UV/IR) singularities, are very dangerous, leading to
a breakdown of most of the perturbative schemes. Even when the UV/IR
infrared singularities are only logarithmic, the mere separation of
contributions into planar (UV divergent) and nonplanar (UV finite but
divergent whenever the external momenta tends to zero) parts,
typical of the Weyl-Moyal method, may lead to inconsistencies in the
renormalization program so that the model under scrutiny becomes
nonrenormalizable. Examples where such situation occurs are provided
by the four dimensional $O(N)$ linear sigma model with $N>2$
\cite{campbell} and the $1/N$ expansion of the $O(N)$ Gross-Neveu (GN)
model in $2+1$ dimensions~\cite{semenoff,gomes}. In both cases the
feature responsible for the failure of the renormalization procedure
is the existence of a parameter whose renormalization in the
commutative setting secures the elimination of the UV divergence of
two different structures.  For the linear sigma model it is the
pion mass counterterm which enforces both the vanishing of the pion mass and the finiteness of the gap equation.
In the GN model the coupling constant renormalization
plays a double role enforcing the gap equation and also eliminating the
UV divergence in the two point vertex function of the auxiliary field
introduced to implement the $1/N$ expansion. It was proved that enlarging
the models, specifically, the gauging of the linear sigma model and
the supersymmetrization of the GN model furnished consistent theories
without the difficulty aforementioned.

In the present work we will investigate an alternative procedure to introduce noncommutativity in field theories aiming at the construction of a consistent GN model  without the necessity of supersymmetrization. More precisely, we will analyze a coherent state representation~\cite{smailagic}, which is constructed such that only the unperturbed propagators are affected by the noncommutativity. As a consequence Feynman diagrams are not separated into planar and nonplanar parts and, in general, all amplitudes are ultraviolet finite (some recent applications of this method are in~\cite{casadio}). We stress that this formalism is unrelated with the ordering prescriptions inherent to the Weyl correspondence in canonical noncommutative field theories, a point which we clarify in the Appendix.

It has been argued also that, under some simple assumptions on the noncommutativity matrix, Lorentz preserving noncommutative models  may  be constructed on even spacetime dimensions, using coherent states. In our case, because the spacetime dimension is odd, Lorentz symmetry is being explicitly broken. However we may envisage the possibility that the breaking occurs only at very high energies so that its net effect is strongly suppressed at our energy scale. We will show that this is indeed the case in this model. One may entertain the hope that the same mechanism may work for more realistic field theories operating in four dimensional spacetime~\cite{ourlv}.

This work is organized as follows. The commutative Gross-Neveu model and its canonical noncommutative counterpart are described in Section~\ref{ncgn}. Using a coherent state approach, in Section~\ref{gncs}, we show that the problems in the canonical approach are circumvented. In this context we describe the main properties of the model and discuss the problem of Lorentz violations. Our conclusions are contained in Section~\ref{conclusions}. The Appendix contains an analysis of the various ordering prescriptions used in the canonical noncommutative field theories, and how they may relate (or not) with our approach.

\section{The Gross-Neveu model and the canonical noncommutativity}
\label{ncgn}

The commutative Gross-Neveu model is specified by the Lagrangian density
\begin{equation}
{\cal L} =\frac{i}2\overline \psi \not \!\partial \psi -\frac{\sigma}2 (\overline \psi
\psi)- \frac{N}{4g}\sigma^2,\label{1}
\end{equation}

\noindent
where $\psi_i$, $i=1,\ldots N$, are two-components Majorana fields and $\sigma$ is an auxiliary field (note that the replacement of  the $\sigma$ field's equation of motion in  Eq. (\ref{1}) leads to the usual four-fermion interaction). At the quantum level, it is convenient to replace $\sigma$ by $\sigma+M$ where $M$ is the vacuum expectation value of the original $\sigma$. The new Lagrangian is
\begin{equation}
{\cal L} =\frac{i}2\overline \psi \not \!\partial \psi - \frac{M}2\overline \psi
\psi-\frac{\sigma}2 (\overline \psi
\psi)- \frac{N}{4g}\sigma^2-\frac{N}{2g} M \sigma.\label{2}
\end{equation}

\noindent
Observe now that the coupling constant renormalization, $1/g \rightarrow 1/g_R + \Delta$,
where $g_R$ is the renormalized coupling constant, equally affects the tadpole and the two point
vertex function of the auxiliary field. In fact, since  the  $\sigma$ field now has zero vacuum expectation value,
the gap equation
\begin{equation}
 \frac{M}{2g}-i\, \int \frac{d^Dk}{(2\pi)^D}\frac{M}{k^{2}-M^2}=0,\label{3}
\end{equation}

\noindent
must be obeyed. Now, the computation of the two point vertex function of the $\sigma$ field
leads to
\begin{eqnarray}
\Gamma_{\sigma}^{(2)}&=&-\frac{iN}{2g}-N\int\frac{d^Dk}{(2\pi)^D}\frac{k\cdot (k+p) + 
M^2}{(k^2-M^2)[(k+p)^2-M^2]}\nonumber\\ 
&& = -\frac{iN}{2g}+ N\int \frac{d^Dk}{(2\pi)^D}\frac{1}{k^{2}-M^2}\nonumber\\
&&+\frac{(p^2-4M^2)N}2\int\frac{d^Dk}{(2\pi)^D} 
\frac{1}{(k^2-M^2)[(k+p)^2-M^2]},\label{4}
\end{eqnarray}

\noindent
which shows that the replacement  $1/g\rightarrow 1/g_R + \Delta$  eliminates divergences both in the gap equation and in the propagator for the auxiliary field.

We now consider  the extension of the above model to a noncommutative space characterized
by the commutation relation between coordinate operators
\begin{equation}
 [\hat q^{\mu},\,\hat q^{\nu}]= i\Theta^{\mu\nu}\label{5}
\end{equation}

\noindent
and set $\Theta^{i0}=0$ to keep time local, thus avoiding unitarity/causality problems~\cite{gomis}. We adopt also the notation  $\Theta^{ij}=i\varepsilon^{ij}\Theta$ where $\varepsilon^{ij}$ is the Levi-Civit\`a anti-symmetrical symbol.

In the Weyl-Moyal approach to noncommutative field theories, the pointwise multiplication of fields is replaced by Moyal product between them. For a given model the propagators are the same as in the corresponding commutative model  but the vertices are modified by trigonometric factors. As a consequence, in our situation the gap equation remains unchanged whereas the factor $\cos^2(k^\mu p^\nu\Theta_{\mu\nu})$ appears in the integral in the first line of Eq.~(\ref{4}), thus leading to
\begin{eqnarray}
\Gamma_{\sigma}^{(2)}&=&-\frac{iN}{2g}+\frac{N}{2} \int \frac{d^Dk}{(2\pi)^D}\frac{1}{k^{2}-M^2}+\nonumber\\
&& \frac{(p^2-4M^2)N}{4}\int\frac{d^Dk}{(2\pi)^D}\frac{1}{(k^2-M^2)[(k+p)^2-M^2]}\nonumber\\
&& +\frac{N}{2} \int \frac{d^Dk}{(2\pi)^D}\frac{\cos(2k \wedge p )}{k^{2}-M^2} \nonumber\\
&& + \frac{(p^2-4M^2)N}{4}\int\frac{d^Dk}{(2\pi)^D}\frac{\cos(2k \wedge p )}{(k^2-M^2)[(k+p)^2-M^2]}\label{51}\,.
\end{eqnarray}

\noindent
The integrals in the second line of the above equation are finite due to the trigonometric factor $\cos(2k \wedge p )$, however, the counterterm $\Delta$, which is fixed by the gap equation, does not eliminate anymore the divergent integral in the first line, because of the $1/2$ factor appearing there. The model has become nonrenormalizable!

\section{The Noncommutative Gross-Neveu model using coherent states}
\label{gncs}

Faced with the problem outlined in the previous section, one could try to find whether alternative approaches to spacetime noncommutativity could be used to modify the Gross-Neveu model without spoiling the delicate equilibrium between the renormalization of the gap equation and the auxiliary field propagator. One possibility is to use the coherent state approach proposed in~\cite{smailagic}. In this case, the commutation relation~(\ref{5}) between the coordinates $\hat q^1 $ and $\hat q^2$ implies that the complex variable $\hat z=\frac{\hat q^1+i\hat q^2}{\sqrt{2}}$ and its complex conjugate $\hat z^\dag$ satisfy
\begin{equation}
 [\hat z,\,\hat z^\dag] =  \Theta.\label{6}
\end{equation}

Defining a ``vacuum'' state through

\begin{equation}
 \hat z|\,0>=0 \qquad \qquad <0\,|\hat z^\dag=0,\label{7}
\end{equation}

\noindent
we may construct eigenstates of the ``number'' operator $\frac{\hat z^\dag \hat z}{\Theta}$ by applying powers of the
``creation'' operator $\hat z^\dag$ to the vacuum,
\begin{equation}
 \frac{\hat z^\dag \hat z}{\Theta} (\hat z^\dag)^n |\,0> = n (\hat z^\dag)^n |\,0>\label{8}
\end{equation}

\noindent
Coherent states, which are eigenstates of the annihilation operator $\hat z$, i.e., $\hat z|\,\alpha>=\alpha|\,\alpha>$, are given by
\begin{equation}
 |\,\alpha> \,=\, \exp\left( -\frac{1}{2} |\alpha|^{2} \right) \exp\left(\alpha \hat z^{\dag}\right)\,|\,0>\label{9} \,.
\end{equation}

\noindent
Introducing commutative coordinates by $\alpha=x+iy$, to each classical field $f(x)$  the Fourier representation
\begin{equation}
 \hat \Phi(\hat q)= \int \frac{d^3 k}{(2\pi)^3} \, {\rm e}^{-ik\hat q}\tilde \psi(k),\label{10}
\end{equation}

\noindent 
where $\tilde \psi(k)$ denotes the Fourier transform of $f(x)$, associates a field operator $\hat \Phi(\hat q)$.  The expectation value of this operator defines a classical field
\begin{equation}
\psi(x)\,=\, <\alpha\,|\hat \Phi(\hat q)|\,\alpha> \,=\, \int \frac{d^3 k}{(2\pi)^3} \, {\rm e}^{-i k x - \frac{1}{4}\Theta |\vec{k}|^2}\tilde \psi(k).\label{11}
\end{equation}

The above expression defines the  coherent representation for the  classical field $f(x)$.  If $f(x)$ is a quantized
free scalar field the propagator for the coherent field $\psi(x)$ is given by
\begin{eqnarray}
\Delta_F(x-y)&\equiv & <0\,|T \psi(x)\,\psi(y)|\,0> \nonumber\\
&=&  \int \frac{d^3 k_1}{(2\pi)^3} \frac{d^3 k_2}{(2\pi)^3}{\rm e}^{-i k_1 x -i k_2 y} \, {\rm e}^{-\frac14 \Theta(|\vec{k}_1|^2 +|\vec{k}_2|^2)} (2\pi)^3 \delta^3 (k_1+k_2) \frac{i}{k^{2}_{1} - m^2}\nonumber\\
&=&\int \frac{d^3 k}{(2\pi)^3} \, {\rm e}^{-i k (x-y)} \frac{i}{k^2 - m^2}{\rm e}^{- \frac{1}{2}\Theta |\vec{k}|^2}\label{12}
\end{eqnarray}

In the sequel we are going to formulate interacting field theories as the quantum versions of the classical fields above defined. In a more precise terms, the Lagrangian density for a self-interacting field $\psi$ reads
\begin{equation}
 {\cal L}(x) = \psi(x) \, {\rm e}^{-\frac{\Theta}{2}\vec{\nabla}^2}\,{\cal O} \, \psi(x) + {\cal L}_{int}(x),
\end{equation}

\noindent
where ${\cal O}= -(\partial_\mu \partial^\mu + m^2)$ or ${\cal O}=(i\not\! \partial -M)$ for scalar of fermionic fields, respectively. The interacting Lagrangian density ${\cal L}_{int}(x)$ is a polynomial in the basic field and its derivatives. Notice that the extra nonlocal factor in the free part of the Lagrangian was devised as to reproduce the result~(\ref{12}). However, one chooses the interacting Lagrangian as some local product of the
basic field $\psi(x)$. 

In the case of the Gross-Neveu model each field is replaced by its corresponding field representative, using the correspondence in Eq.~(\ref{10}), so that the interaction vertices looks the same as in the commutative situation. The computation of the gap equation now leads to 
\begin{equation}
 \frac{1}{2g}-i\,\int \frac{d^3k}{(2\pi)^3}\frac{1}{k^{2}-M^2}{\rm e}^{- \frac{1}{2}\Theta |\vec{k}|^2}\label{13}=0.
\end{equation}

\noindent
and we obtain
\begin{equation}
\frac{1}{g}=  \frac{{\rm e}^{\frac{\Theta}{2}M^2}}{2\sqrt{2\pi\Theta}} {\rm Erfc}\left[M\sqrt{\frac{\Theta}{2}}\right] =
\frac{1}{2\sqrt{2\pi\Theta}}- \frac{M}{2\pi} + \mathcal{O}(\Theta)\label{14}
\end{equation}

\noindent
where ${\rm Erfc}[z]=\frac{2}{\sqrt{\pi}}\int_{z}^{\infty} {\rm e}^{-t^2} dt $ denotes the complementary error function and the last equality indicates the leading behavior of the left hand side for small $\Theta$. The gap equation~(\ref{13}) is finite for non-zero $\Theta$, and can be made regular in the $\Theta \rightarrow 0$ limit by means of a coupling constant renormalization $\frac{1}{g} \rightarrow \frac{1}{g_R}-\frac{1}{2\sqrt{2\pi\Theta}}$.

Let us now consider the  propagator for the auxiliary field $\sigma$, $\Delta_\sigma=-[\Gamma^{(2)}_{\sigma}]^{-1}$, where
\begin{equation}
 \Gamma^{(2)}_{\sigma}(p)= \frac{iN}{2g}{\rm e}^{\frac{\Theta}{2}|\vec p|^2}- \Sigma_{\sigma}(p)\label{15}
\end{equation}

\noindent
and
\begin{eqnarray}
\label{eq18}
 \Sigma_{\sigma}(p) &=& -N \int\frac{d^3k}{(2\pi)^3}\frac{k\cdot (k+p) + 
M^2}{(k^2-M^2)[(k+p)^2-M^2]}  {\rm e}^{-\frac{\Theta}{2}|\vec k|^2}{\rm e}^{-\frac{\Theta}{2}|\vec k+\vec p|^2}
\label{16} \\
&=& -N \int\frac{d^3k}{(2\pi)^3}\frac{{\rm e}^{-\frac{\Theta}{2}|\vec k|^2}{\rm e}^{-\frac{\Theta}{2}|\vec k+\vec p|^2}}{k^2-M^2} \nonumber\\
&&+ \frac{(p^2 - 4 M^2)N}{2}\int\frac{d^3k}{(2\pi)^3}\frac{{\rm e}^{-\frac{\Theta}{2}|\vec k|^2}{\rm e}^{-\frac{\Theta}{2}|\vec k+\vec p|^2}}{(k^2-M^2)[(k+p)^2-M^2]} \nonumber
\end{eqnarray}

\noindent
Notice that all integrals are finite, the integrands being exponentially damped as the loop momenta increases. 
{In this formalism, the noncommutativity of spacetime manifests itself in the appearance of an effective regularization of the loop integrals. It must be stressed, however, that the interpretation of~(\ref{16}) is not that of a regularized Feynman integral, as in usual quantum field theory, since the ``cutoff'' $1/\Theta$ is a natural scale which is not introduced as an intermediate step in the renormalization procedure. In this context, the scale $\Theta$ is in principle small, but \textit{finite}. 

We may be interested in studying the commutative limit of our model, however, we find that it is not possible to get a smooth $\Theta \rightarrow 0$ limit. This is so because the leading behavior for small $\Theta$ in Eq.~(\ref{eq18}) is different from the one in Eq.~(\ref{14}). As in the canonical noncommutativity approach, the delicate equilibrium between the gap equation and the auxiliary field propagator renormalizations is lost but, at least, here the problem only appears if one insists in having a smooth $\Theta \rightarrow 0$ limit. Whenever the noncommutativity parameter is kept finite, the coherent states approach is able to keep all divergences under control. Also, despite the non-analiticity in $\Theta$, there is no UV/IR mixing present in this approach, since all integrals are regular for vanishing external momentum. With the coupling constant renormalization previously adopted, $\Delta_{\sigma} \rightarrow 0$ leading to a peculiar theory as $\Theta \rightarrow 0$.

Alternatively, one could use the coupling constant renormalization to eliminate the divergent integral in Eq.~(\ref{16}), at the price of a $1/\sqrt{\Theta}$ singularity in the gap equation~(\ref{13}), implying in the vanishing of the renormalized coupling constant in the $\Theta \rightarrow 0$ limit (characterizing an asymptotically free model). Again, this is in contrast with the UV/IR problem in canonical noncommutative theories, which threatens the renormalization program because of the blow up of higher order quantum corrections~\cite{minwalla}. 

We turn our attention now  to the $\psi$ field  two point function, whose leading correction is given by
\begin{equation}
 \Sigma (p) = \int \frac{d^3 k}{(2\pi)^3} \frac{\not \! \! k+ \not \! \! p+M}{(k+p)^2- M^2}\Delta_\sigma(k)
{\rm e}^{-\frac{\Theta}{2}|\vec k+\vec p|^2}\label{18}
\end{equation}

\noindent
This expression is clearly well defined as far as $\Theta \neq 0$. Nonetheless, it linearly diverges as $\Theta\rightarrow 0$ so that, to get a smooth limit, we can renormalize the model by imposing that the $\psi$ field propagator satisfies
\begin{equation}
 \left[ \Delta_\psi(p)\right]^{-1} {\buildrel {p \rightarrow 0} \over \approx } - i  \left( \not \! p - m  \right) + \mathcal{O}(p^2)
\label{19}\,.
\end{equation}

Another aspect that raises concern is the Lorentz violation (LV) embodied in the commutation
relation~(\ref{5}). Indeed, several authors have pointed out the difficulties
in conciliating the LV induced in canonical noncommutative field theories
with the known experimental constraints~\cite{NCLV,collins,ourlv}, and this
have motivated the search for Lorentz-preserving noncommutative models~\cite{carlson}.

As far as the coherent states approach is concerned, it was claimed that in even dimensional spacetime it is possible to avoid the LV by a clever choice of the noncommutativity matrix $\Theta_{\mu\nu}$~\cite{smailagic}. In opposition to this result, in odd spacetime dimensions as it is our case, the use of the coherent state basis inevitably leads to a LV. However, one may argue that, if the Green functions of the basic field $\psi$ can be made analytical in $\Theta$, the breaking is necessarily small for small $\Theta$. We can explicitly check this for the two-point function in Eq.~(\ref{18}).

From a theoretical standpoint, the parameter
\begin{equation}
\xi\left[\Pi\left(p\right)\right]\,=\,\left[\left(\frac{\partial^{2}}{\partial\left(p^{0}\right)^{2}}+\frac{\partial^{2}}{\partial\left(p^{1}\right)^{2}}\right)\Pi\left(p\right)\right]_{p=0}\,,
\end{equation}

\noindent
suggested in~\cite{collins} was used to measure the LV in the scalar amplitude $\Pi\left(p\right)$. That $\xi$ is an adequate measure follows from the fact that it always vanishes if $\Pi\left(p\right)$ is Lorentz-invariant, while $\xi$ differs from zero in the Lorentz violating case ($\xi$ corresponds to a Lorentz-violating correction to the dispersion relation of the scalar particle). 

As the basic field of the Gross-Neveu model is a spinor, some modification is necessary and we propose the use of
\begin{equation}
\chi\left[\Sigma\left(p\right)\right]\,=\,\left[\left(\frac{\partial}{\partial\left(p^{0}\right)}+\frac{1}{2}\sum_{i=1}^{2}
\gamma_{i}\frac{\partial}{\partial\left(p^{i}\right)}\right)\Sigma\left(p\right)\right]_{p=0}\,.\label{eq:ki}
\end{equation}

\noindent
as a measure of the LV in the fermion self-energy $\Sigma\left(p\right)$. One can check that $\chi=0$ in a Lorentz invariant theory. By applying this differential operator to (\ref{18}) we obtain
\begin{equation}
\chi\,=\,\chi^{\left(0\right)}+\Theta\chi^{\left(1\right)}\,,\label{eq:kign}
\end{equation}

\noindent
where
\begin{equation}
\chi^{\left(0\right)}\,=\,-2\int\frac{d^{3}k}{\left(2\pi\right)^{3}}\left(\gamma^{0}k^{0}+\frac{1}{2}\sum_{i=1}^{2}\gamma^{i}k^{i}\right)\frac{\not  \! k+M}{\left(k^{2}-M^{2}\right)^{2}}\Delta_{\sigma}\left(k\right)e^{-\frac{\Theta}{2}|\vec{k}^{2}|}
\end{equation}

\noindent
and
\begin{equation}
\chi^{\left(1\right)}\,=\,\frac{1}{2}\int\frac{d^{3}k}{\left(2\pi\right)^{3}}\left(\sum_{i=1}^{2}\gamma^{i}k^{i}\right)\frac{\not \! k+M}{k^{2}-M^{2}}\Delta_{\sigma}\left(k\right)e^{-\frac{\Theta}{2}|\vec{k}^{2}|}\,.
\end{equation}

Habitually, we conjecture that $\Theta$ is very small, being of the order of two powers of the Planck length. From this perspective, $\chi^{\left(1\right)}$ is a very small effect of the LV, but the presence of the first term, $\chi^{\left(0\right)}$, may appear at first sight troublesome. Such worries are unfounded since, to enforce the renormalization condition~(\ref{19}), one has to replace  $\Sigma(p)$ by 
\begin{equation}
 \Sigma_{R}\left(p\right)\,=\,\Sigma\left(p\right)-\Sigma\left(0\right)-p^{\mu}\left[\frac{\partial}{\partial\left(p^{\mu}\right)}\Sigma\left(p\right)\right]_{p=0}\,
\end{equation}

\noindent
and it is easily found that $\chi\left[\Sigma_R \left(p\right)\right]\,=\,0$ so that a large Lorentz violation does not appear. We would like to stress that, in the canonical approach to noncommutativity, i.e., by use of the Moyal product, the above procedure is not available as the planar parts of Feynman amplitudes are in general not renormalizable. This is a clear advantage of the coherent state approach.

\section{Conclusions}
\label{conclusions}

In this paper, we shown that the use of a coherent states approach for the introduction of spacetime noncommutativity avoids serious problems with the canonical noncommutative extension of the Gross-Neveu model. In this last context, the fact that part of the original ultraviolet divergences survive, and that a single coupling constant renormalization is available to eliminate divergences in two very different structures, makes the theory non-renormalizable. However, in the coherent states formalism, one evades such troubles, and the resulting noncommutative Gross-Neveu model is finite and free of UV/IR mixing for non-vanishing noncommutativity parameter $\Theta$. 

We have also studied the generation of Lorentz violating corrections to the dispersion relation of the model. For finite $\Theta$, if we do not perform any subtraction on the Green functions, large Lorentz violation do appear. This problem can be surmounted if we insist that our model should have a well-behaved $\Theta \rightarrow 0$ limit. In this case, a renormalization procedure must be implemented, and this takes care of the LV. Curiously, the $\Theta \rightarrow 0$ limit is not the commutative Gross-Neveu model, but an asymptotically free theory. 

In this work, we adopted the idea that the ``blurring'' effect of the noncommutativity
of coordinates, which would be induced by quantum gravity~\cite{dopli},
is completely embodied in the Fourier transform of a single field, as
in Eq.~(\ref{11}). This induces the modified propagators that we used. As for the interaction part, we choose the simplest possibility, which is a local product of the classical field defined by Eq.~(\ref{11}).
There are certainly more complicated choices, yet even this simple
possibility have interesting implications, as our analysis of the
Gross-Neveu model indicates.

\vspace{1cm}

{\bf Acknowledgements.} This work was partially supported by Funda\c{c}\~{a}o de Amparo 
\`{a} Pesquisa do Estado de S\~{a}o Paulo (FAPESP), Conselho Nacional de Desenvolvimento Cient\'{\i}fico e Tecnol\'{o}gico (CNPq) and Coordena\c{c}\~{a}o de Aperfei\c{c}oamento de Pessoal de N\'{\i}vel Superior (CAPES). The work of A. F. F. was supported by FAPESP,  project 04/13314-4.

\appendix

\section{On the question of the ordering prescription}
\label{ordering}

The canonical noncommutativity approach is based on a correspondence between classical functions and quantum operators, which fixes the form of the Moyal product used to define noncommutative models. This correspondence is defined up to an arbitrary ordering prescription. To take this into account, we introduce a generalized
Weyl correspondence~\cite{wolf}, 
\begin{equation}
\hat{\Phi}^{\left(f\right)}\left(\hat{q}\right)\,\equiv\,\int d^{2}x\,\phi\left(x\right)\Delta^{\left(f\right)}\left(\hat{q}-x\right)\,,\label{30}
\end{equation}

\noindent
where
\begin{equation}
\Delta^{\left(f\right)}\left(\hat{q}-x\right)\,=\,\int\frac{d^{2}k}{\left(2\pi\right)^{2}}\, f\left(k\right)e^{-ik\left(\hat{q}-x\right)}\,,\label{31}
\end{equation}

\noindent
and $f\left(k\right)=f\left(k_{1},k_{2}\right)$ is an arbitrary function encoding the ordering
ambiguity in the relation between functions $\phi\left(x\right)$
and operators $\hat{\Phi}^{\left(f\right)}\left(\hat{q}\right)$. One should only requires that $f$ nowhere vanishes and that $f(0)=1$. Popular ordering choices, like normal ordering, Weyl ordering and
so on, can be implemented by particular choices of $f\left(k\right)$; in particular, $f\left(k\right)=1$ yields the usual Weyl correspondence. 

The inverse correspondence is given by
\begin{equation}
\phi\left(x\right)\,=\,\mbox{Tr}\left[\hat{\Phi}^{\left(f\right)}\left(\hat{q}\right)\Delta^{\left(\tilde{f}\right)}\left(\hat{q}-x\right)\right]\,,\label{32}
\end{equation}

\noindent
where $\tilde{f}\left(k\right)=1/f\left(-k\right)$. Here, the trace
is normalized as $\mbox{Tr}\left(e^{-ik\hat{q}}\right)=\left(2\pi\right)^{2}\delta^{2}\left(k\right)$.
From this inverse map, one defines a star-product involving
$n$ classical functions,
\begin{eqnarray}
\phi_{1}\left(x\right)\star && \phi_{2}\left(x\right)\star\cdots\star\phi_{n}\left(x\right)\,\equiv\,\mbox{Tr}\left[\hat{\Phi}_{1}^{\left(f\right)}\left(\hat{q}\right)\hat{\Phi}_{2}^{\left(f\right)}\left(\hat{q}\right)\cdots\hat{\Phi}_{n}^{\left(f\right)}\left(\hat{q}\right)\Delta^{\left(\tilde{f}\right)}\left(\hat{q}-x\right)\right]\nonumber \\
=\,\int && \left[\prod_{i=1}^{n}\frac{d^{2}k_{i}}{\left(2\pi\right)^{2}}\right]\left[\prod f\left(k_{i}\right)\right]\tilde{f}\left(-\sum k_{i}\right)e^{-\frac{i}{2}\sum_{i<j}k_{i}\wedge k_{j}}\nonumber \\
 && \times e^{-i\left(\sum_{i}k_{i}\right)x}\tilde{\phi}_{1}\left(k_{1}\right)\tilde{\phi}_{2}\left(k_{2}\right)\cdots\tilde{\phi}_{n}\left(k_{n}\right)\,,\label{33}
\end{eqnarray}

\noindent
where $\tilde{\phi}_{i}\left(k_{i}\right)$ is the Fourier transform
of $\phi_{i}\left(x\right)$. Note that all integrals are two-dimensional
since there are two noncommuting coordinates $\hat{q}_{1}$ and $\hat{q}_{2}$,
time being a commutative parameter untouched by the correspondence.
Thus, the space-time integral of~(\ref{33}) can be cast as
\begin{eqnarray}
&& \int d^{3}x \phi_{1}\left(x\right)\star  \phi_{2}\left(x\right)\star\cdots\star\phi_{n}\left(x\right)\,=\,\int\left[\prod_{i=1}^{n}\frac{d^{3}k_{i}}{\left(2\pi\right)^{3}}\right]\left(2\pi\right)^{3}\delta^{3}\left(\sum_{i}k_{i}\right)\nonumber \\
&& \,\, \times f\left(k_{1}\right)f\left(k_{2}\right)\cdots f\left(k_{n}\right)e^{-\frac{i}{2}\sum_{i<j}k_{i}\wedge k_{j}}\,\tilde{\phi}_{1}\left(k_{1}\right)\tilde{\phi}_{2}\left(k_{2}\right)\cdots\tilde{\phi}_{n}\left(k_{n}\right)\,.\label{34}
\end{eqnarray}

As becomes clear from Eq.~(\ref{33}), the usual Moyal product is
obtained for all $f$'s satisfying $f\left(k+q\right)=f\left(k\right)f\left(q\right)$,
which obviously happens for $f=1$ but not for other popular orderings,
like for example the normal ordering, which is reproduced by 
\begin{equation}
f\left(k\right)=\exp\left(\frac{\theta}{2}|\vec{k}^{2}|\right)\,.\label{35}
\end{equation}

\noindent
Typically, both propagators and vertices will be modified by the $f$
factors present in~(\ref{34}), but these modifications will disappear
when one calculates the quantum corrections to the effective action
of the theory, and the result will be the same as the one in the usual
Moyal-product approach. Indeed, for the quadratic part of the action,
one has
\begin{equation}
\int d^{3}x\,\phi_{1}\left(x\right)\star\mathcal{O}\phi_{1}\left(x\right)=\int\frac{d^{3}k}{\left(2\pi\right)^{3}}f\left(k\right)f\left(-k\right)\tilde{\phi}_{1}\left(k_{1}\right)\tilde{\mathcal{O}}\tilde{\phi}_{2}\left(k_{2}\right)\,,
\label{36}
\end{equation}

\noindent
so that internal propagators acquire a $1/f^{2}$ factor. However,
this $1/f^{2}$ cancel the $f$'s arising from the vertices attached
to the ends of the internal lines. 
In this way, even if the Moyal
product is sensible to the ordering choice, the quantum theory seems to be
actually independent of the operator ordering~\cite{chaichian,hammou}. 
One could allow for different orderings for the free and interactions parts of the Lagrangian~\cite{chaichian} but, even in this case, the ``standard'' Moyal factor $e^{-\frac{i}{2}\sum_{i<j}k_{i}\wedge k_{j}}$ would be present. 
This is essentially different from the coherent states approach where the damping exponentials in the free propagators are not canceled in the computation of general Feynman amplitudes, and the ``standard'' Moyal factor is absent. 
This happens because the noncommutativity is already embodied in the Fourier transform of a single field, as explicitly shown in Eq.~(\ref{11}), and the interaction Lagrangian is chosen to be a simple, local, product of fields.

\end{document}